\documentclass[twocolumn,secnumarabic,amssymb,amsfonts,nobibnotes,aps,jcp,showpacs]
{revtex4-1}

\usepackage{amsmath}
\usepackage{latexsym}
\usepackage {mathrsfs}
\usepackage{float}
\usepackage{amssymb}
\usepackage{amsfonts}
\usepackage{graphicx}
\usepackage{textcomp}
\usepackage{hyperref}
\usepackage {mathrsfs}
\usepackage {pifont}
 1

\newcommand{\dummy}

\begin{document}
\title{Nucleation and Growth of Droplets in Vapor-Liquid Transition}
\author{Sutapa Roy and Subir K. Das$^*$}
\affiliation{Theoretical Sciences Unit, Jawaharlal Nehru Centre for Advanced 
Scientific Research, Jakkur P.O, Bangalore 560064, India}
\date{\today}

\begin{abstract}
Results for the kinetics of vapor-liquid transitions, following temperature 
quenches with different densities, are presented from the molecular dynamics 
simulations of a Lennard-Jones system. For critical density, bicontinuous 
liquid and vapor domains are observed which grow with time obeying the 
prediction of hydrodynamic mechanism. On the other hand, for quenches with 
density significantly below the critical one, phase separation progresses 
via nucleation and growth of liquid droplets. In the latter case, Brownian 
diffusion and collision mechanism for the droplet growth is confirmed. 
We also discuss the possibility of inter-droplet interaction leading to a 
different amplitude in the growth law. Arguments for faster growth, observed 
at early time, are also provided.
\end{abstract}
\maketitle

\par
\hspace{0.2cm} The subject of nucleation and growth is of significant 
importance in many branches of physics, chemistry and engineering 
\cite{wadhawan,jones}. In spite of that, answers to many fundamental questions 
in this area still remain challenging \cite{wadhawan}. The phenomena of growth 
is understood in some detail in simple situations, e.g., late time 
non-equilibrium dynamics in bulk solid mixtures with critical (symmetric) 
compositions (which belongs to the so called category of spinodal 
decomposition), via simple Ising model or Cahn-Hilliard equation 
\cite{wadhawan,binderbook,bray,onuki,jones}. In this case one obtains 
interconnected domain morphology. Even for such simple situations our 
knowledge appears rather incomplete when one considers dynamics at early time 
\cite{suman1,suman2} or when one puts the systems in confinement 
\cite{das1,das2}. Both these examples are related to nanoscopic length scales. 
The situation is far worse in fluids where one encounters greater complexity 
due to the influence of hydrodynamics \cite{siggia,furukawa1,furukawa2}. In 
this work we address the problem of vapor-liquid phase separation in extreme 
off-critical situation that gives rise to droplet morphology and thus related 
to the nucleation phenomena. This, of course, has direct relevance in 
nano-science and technology. For the ease of a precise definition of the 
problem, below we give a brief introduction to the field in the context of 
a binary mixture $(A+B)$. 
\par
\hspace{0.2cm} When a homogeneously mixed system is quenched inside the 
coexistence curve, the system phase separates via formation and growth of 
$A-$rich and $B-$rich domains. Typically, this phase separation is a 
self-similar phenomena \cite{wadhawan,binderbook,bray}, viz., the 
morphology at different times ($t$) are similar except for a change of length 
scale, $\ell(t)$, which is the average size of domains. This fact is reflected 
in the scaling behavior of functions that characterize the pattern formation. 
E.g., the two-point equal time correlation function $C(r,t)$ ($r$ being the 
scalar distance between two points) exhibit the scaling form \cite{bray} 
$C(r,t) \equiv \tilde {C}(r/\ell(t)).$
It has remained a challenge \cite{wadhawan,bray} to obtain the analytical 
form for $\tilde {C}$ when the order-parameter is a conserved quantity, 
as it is in the present context. The growth of $\ell(t)$ typically follows a 
power-law \cite{bray} 
\begin {eqnarray}\label{powerlaw}
 \ell(t)\sim t^{\alpha},
\end{eqnarray}
where the exponent $\alpha$ depends upon the system and order-parameter 
dimensionality, conservation of order-parameter as well as hydrodynamic 
effects. Here we confine ourselves to conserved scalar order-parameter in 
space dimensionality $d=3$. 
\par
\hspace{0.2cm} For diffusive transport, which is true for the entire growth 
dynamics in solid binary mixtures, the rate of change of $\ell(t)$ is related 
to the chemical potential $(\mu)$ gradient as \cite{bray} 
\begin {eqnarray}\label{LS}
\frac{d\ell(t)}{dt} \sim {\frac {1}{\ell(t)}}.\mu={\frac {1}{\ell(t)}}.
{\frac {\gamma}{\ell(t)}},
\end{eqnarray}
where $\gamma$ is the interfacial tension. The solution of Eq.(\ref{LS}) 
provides $\alpha=1/3$. The original derivation due to Lifshitz and Slyozov 
(LS) \cite{lifshitz}, obtained for off-critical situation, is much more 
involved. However, as can be judged from the general nature of the derivation 
in Eq.(\ref{LS}), $\alpha=1/3$ is expected to hold for compositions critical 
as well as off-critical. On the other hand, the hydrodynamic effects cause a 
faster growth at late times for fluids as well as polymers. Typically, 
for a critical quench in a fluid binary mixture one expects three distinct 
regimes of domain coarsening \cite{siggia,furukawa1,furukawa2}, viz., 
diffusive, viscous hydrodynamic and inertial hydrodynamic, with exponents 
$1/3$, $1$ and $2/3$, respectively. Essentially, at late time the 
tube-like interconnected structure facilitates advective transport in fluids. 
However, this picture is not true when one has disconnected droplet morphology 
which geometrically has to be the case for an off-critical quench, if 
the principle of interfacial free energy minimization is accepted. 
\par
\hspace{0.2cm} In the off-critical situation, Binder and Stauffer (BS) 
\cite{binder1,binder2} proposed a Brownian droplet diffusion and collision 
mechanism. There, the time dependence of $\ell(t)$ can be obtained from 
\cite{siggia} ($C$ being a constant)
\begin {eqnarray}\label{BS}
\frac{dn}{dt}=CD{\ell}{n^2},
\end{eqnarray}
where the droplet density $n \propto {\phi/\ell^3}$, $\phi$ being the 
volume fraction of the minority species and $D$ is the droplet diffusion 
constant. Treating $D\ell$ as a constant (according to 
Stokes-Einstein-Sutherland relation \cite{hansen}), from (\ref{BS}) one 
obtains $\alpha=1/3$ \cite{siggia}, same as the LS value. It has been 
pointed out that the ratio of amplitudes $A_{BS}$ and $A_{LS}$, 
in the BS and LS cases, respectively, is 
\cite{siggia,tanaka1,tanaka2}
\begin {eqnarray}\label{ratio}
A_{BS}/A_{LS}=K\phi^{1/3};~K\simeq 6.
\end{eqnarray}
Possibility for $K \simeq 4.84$ has also been argued \cite{tanaka1,tanaka2}. 
It is claimed that the BS scenario will be valid only in the low droplet 
density, for $\phi < 0.06$ \cite{tanaka1,tanaka2}. For high droplet density, 
inter-droplet interaction mechanism, due to concentration gradient, may be 
important \cite{tanaka1,tanaka2,kumaran1,kumaran2}. This latter mechanism, 
though leads to the same exponent, gives amplitude higher than the BS value.
\par
\hspace{0.2cm} Even though we confined our discussion to binary liquids, 
all the above pictures, we believe, should apply to vapor-liquid phase 
separation as well. For off-critical case, this was, in fact, confirmed by 
an experimental study \cite{perrot}. On the other hand, for critical quenches 
the effect of hydrodynamics was observed in molecular dynamics (MD) 
simulations of both \cite{sutapa} liquid-liquid \cite{shaista} and 
vapor-liquid \cite{suman3} phase separations. However, to the best of our 
knowledge, there exists no such computational study with atomistic models 
to verify the predictions in the off-critical case, be it a vapor-liquid 
transition or a liquid-liquid one. In this work, we present extensive results 
from MD simulations \cite{frenkel} to address this important issue of 
nucleation and growth. We confirm that $\alpha=1/3$, demonstrate the Brownian 
motion of droplets, present results related to Eq.(\ref{ratio}) and 
provide arguments for early time fast dynamics.
\par
\hspace{0.2cm} We use a model where particles of equal mass ($m$) at 
positions ${\vec r}_i$ and ${\vec r}_j$ interact, for $r<r_c$, via \cite{suman3} 
$u(r=|{{\vec r}_i}-{{\vec r}_j}|)=U(r)-U(r_c)-(r-r_c)({{dU}/{dr}})_{{r}=r_c},~$
with $U(r)$ being the standard Lennard-Jones (LJ) pair potential with 
inter-particle interaction strength $\varepsilon$. The cut-off distance 
$r_c(=2.5\sigma$, $\sigma$ being the particle diameter$)$ was introduced to 
facilitate faster computation. We obtained the values for $T_c$ and $\rho_c$, 
the latter being the critical value of the density ($\rho=N\sigma^3/L^3$), to be 
approximately $0.9\varepsilon/k_B$ and $0.3$.   
\begin{figure}[htb]
\centering
\includegraphics*[width=0.35\textwidth]{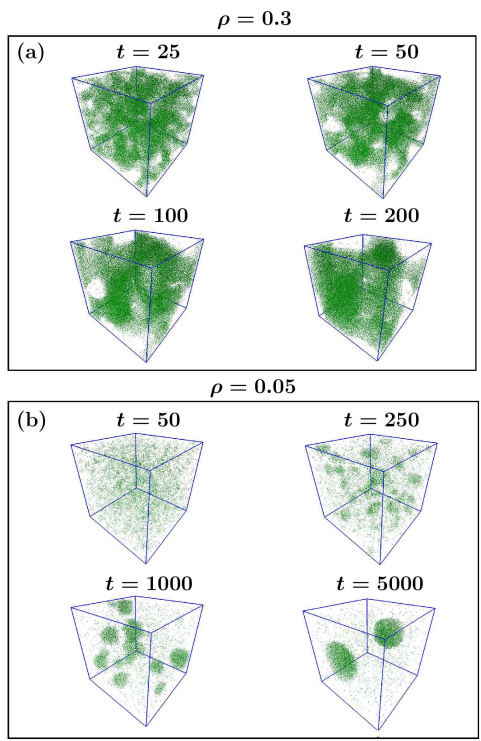}
\caption{\label{fig1}(a) Snapshots from the evolution of the single component 
Lennard-Jones system that exhibits vapor-liquid phase transition. An initial 
configuration with homogeneous density, prepared at a temperature far above 
the critical value with the overall density $\rho=0.3$, was quenched to the 
temperature $0.6$. The linear dimension of the cubic box is $L=64$. The dots 
represent location of particles. (b) Same as (a) but for $\rho=0.05$.}
\end{figure}
We use following units for various relevant quantities. Lengths are expressed 
in units of $\sigma$, temperature in ${\varepsilon}/{k_B}$ and time in 
${m \sigma^2}/\varepsilon$. For the sake of convenience we set 
$\sigma=1$, $\varepsilon=1$, $k_B=1$ and $m=1$. Homogeneous systems with 
different overall densities were prepared at very high temperatures before 
quenching them to $T=0.6$, inside the coexistence curve. Unless otherwise 
mentioned, a Nos\'{e}-Hoover thermostat (NHT) \cite{frenkel}, known for its 
ability to preserve hydrodynamics, was implemented to control the temperature 
in the MD simulations that used integration time step $\Delta t=0.005$. 
Periodic boundary conditions were applied in all directions.
\par
\hspace{0.2cm} In Fig.\ref{fig1}(a) we show the evolution snapshots for 
$\rho=0.3$. It is seen that the phase separation started at a very early 
time, as expected for spinodal decomposition, 
and the domain structures are interconnected. Since the results for 
the hydrodynamic effects on the growth of $\ell(t)$ for this density, 
though at a slightly different temperature, was already presented elsewhere 
\cite{suman3}, for the sake of brevity, we avoid it here. Next we focus on 
the snapshots for $\rho=0.05$, shown in Fig.\ref{fig1}(b). Note that the binodal 
density for the vapor branch at this temperature is $\simeq 0.01$. 
Considering that, we are well inside the metastable region. So, here the 
phase separation progresses via nucleation and growth of droplets. Compared 
to Fig.\ref{fig1}(a), where it is a spontaneous phase separation, the 
nucleation of droplets in Fig.\ref{fig1}(b) is significantly delayed due to 
less super saturation. Our objective in this work is to study the time 
dependence of the growth of these droplets once they are formed.

\begin{figure}[htb]
\centering
\includegraphics*[width=0.35\textwidth]{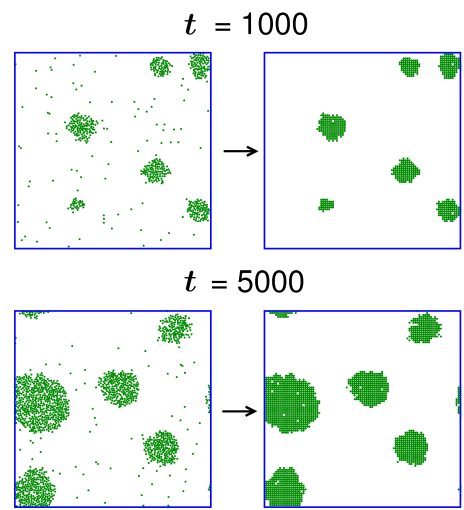}
\caption{\label{fig2}The left panels represent two dimensional cross-sections 
of the evolution snapshots at two different times. Here $\rho=0.05,~L=100$ and 
$T=0.6$. The right panels correspond to the corresponding pictures after 
mapping onto a simple cubic lattice following a method described in the text.}
\end{figure}

\par
\hspace{0.2cm} It is indeed difficult to calculate the droplet radius 
$\ell(t)$ from the continuum configurations seen in Fig.\ref{fig1}. To 
facilitate such calculation, in Fig.\ref{fig2} we describe a simple 
procedure \cite{suman1}. The left panels on this figure are $2-$d slices of 
the snapshots for $\rho=0.05$. The right panels, which look very similar to 
the original ones, are corresponding mapped configurations where the 
particles are moved to the nearest sites of an $L^3$ simple cubic lattice. 
Further, in these mapped configurations, all sites around which the density 
is higher than the critical number have been assigned a spin value $S_i=+1$ 
and the rest got $-1$. Then the collections of up spins constitute liquid 
droplets. Essentially, we are left with a two component Ising model for 
which $C(r,t)$ can be calculated as 
$C(r,t)=\langle S_iS_j\rangle~-~\langle S_i\rangle\langle S_j\rangle;~~r=|i-j|.$
From the number of $+1$ spins in the system one can calculate the fraction 
of volume occupied by liquid droplets. In analogy with a symmetric binary 
liquid, this corresponds to $\phi$ for which we obtain a value $\simeq 0.054$. 
This provides $A_{BS}/A_{LS} \simeq 2.3$.
\begin{figure}[htb]
\centering
\includegraphics*[width=0.38\textwidth]{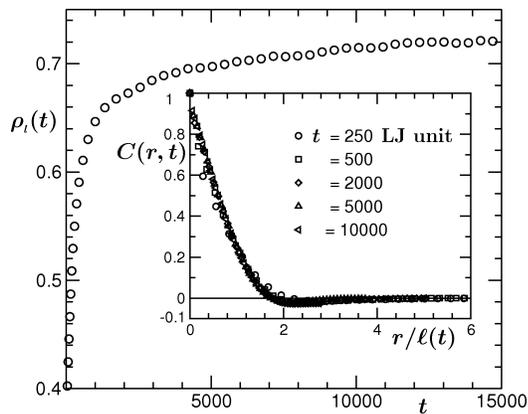}
\caption{\label{fig3}Plot of the density, $\rho_l$, inside the liquid 
droplets as a function of time. Inset: Scaling plot of the correlation 
function $C(r,t)$ as a function of $r/\ell(t)$. Data from five different 
times are used. The values of $\ell(t)$ were obtained from the decay of 
$C(r,t)$ to $1/4$th its maximum value. The results correspond to 
$\rho=0.05$, $L=100$ and $T=0.6$ and on averaging over $10$ independent 
initial configurations.}
\end{figure}

\begin{figure}[htb]
\centering
\includegraphics*[width=0.38\textwidth]{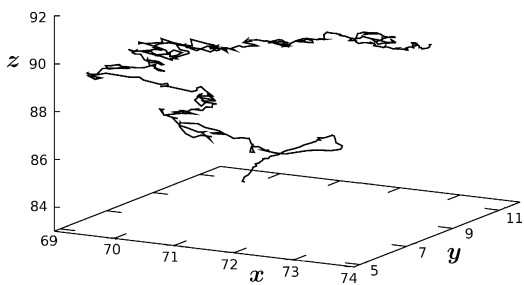}
\caption{\label{fig4}Trajectory of the centre of mass of a droplet in the 
NHT-MD. Only a part of the box is shown, for clarity.} 
\end{figure}

\par
\hspace{0.2cm} From Fig.\ref{fig1}(b) it is quite clear that the densities 
in the liquid and vapor domains take significantly long time to equilibrate. 
For a quantitative picture we have plotted the liquid domain density, 
$\rho_l$, in the main frame of Fig.\ref{fig3} as a function of time. 
The inset of this figure shows the scaling plots of $C(r,t)$ vs $r/\ell(t)$ 
for five different times. The value of $\ell(t)$ was obtained from the 
distance at which $C(r,t)$ decays to $1/{4^{th}}$ its maximum value. 
Starting from $t=2000$ onwards, the data collapse is excellent. The poor 
scaling at early time could be appreciated from the fact that during this 
period $\rho_l$ (and so $\rho_v$) is changing very fast. Note that in 
addition to computing $\ell(t)$ from the decay of $C(r,t)$, we have obtained 
it directly as well. In this direct method \cite{suman1} one sweeps through 
the whole system in different directions to find out number of domains at 
different sizes from which the average value can be calculated in a 
straight forward manner.

\begin{figure}[htb]
\centering
\includegraphics*[width=0.35\textwidth]{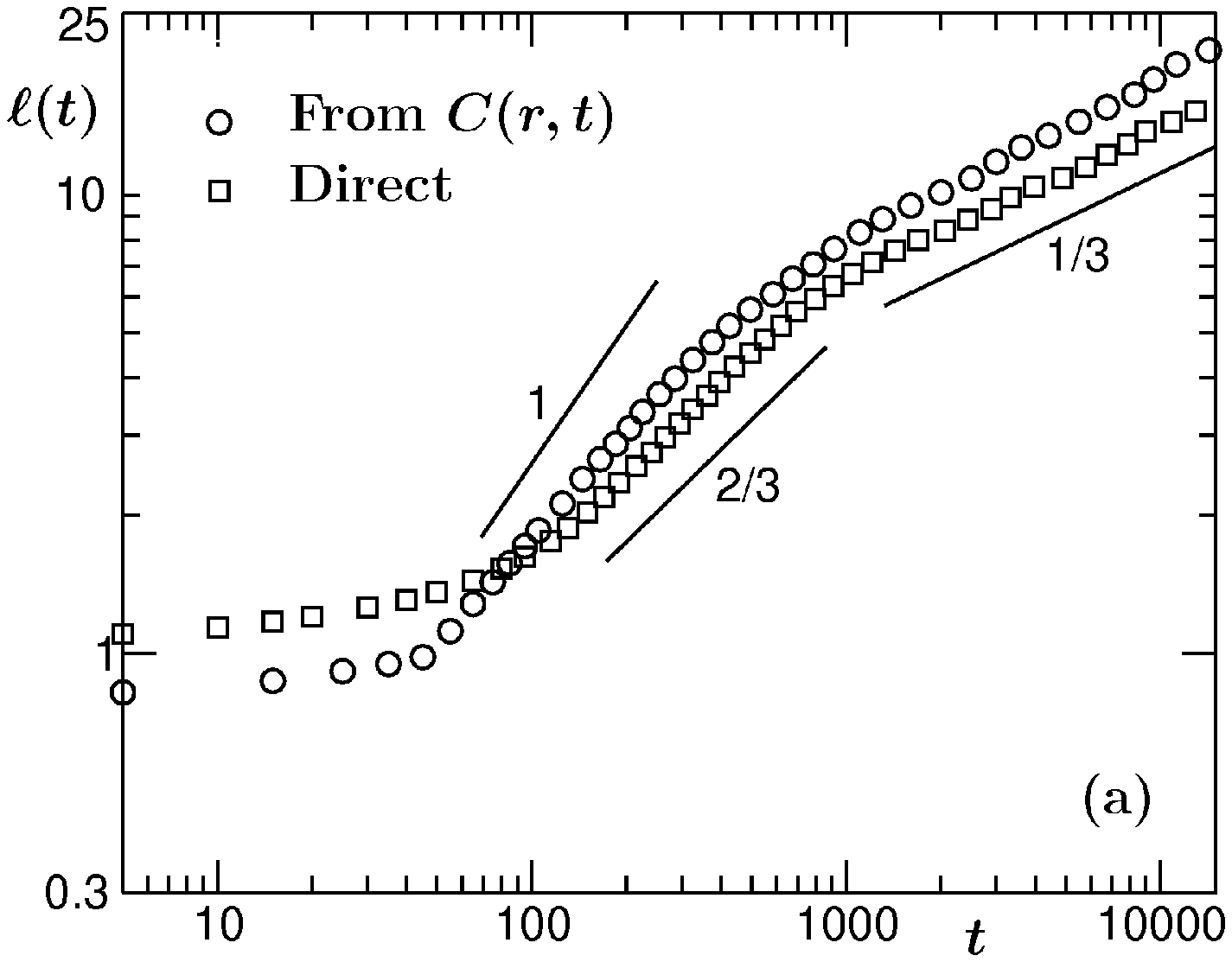}
\vskip 0.2cm
\includegraphics*[width=0.37\textwidth]{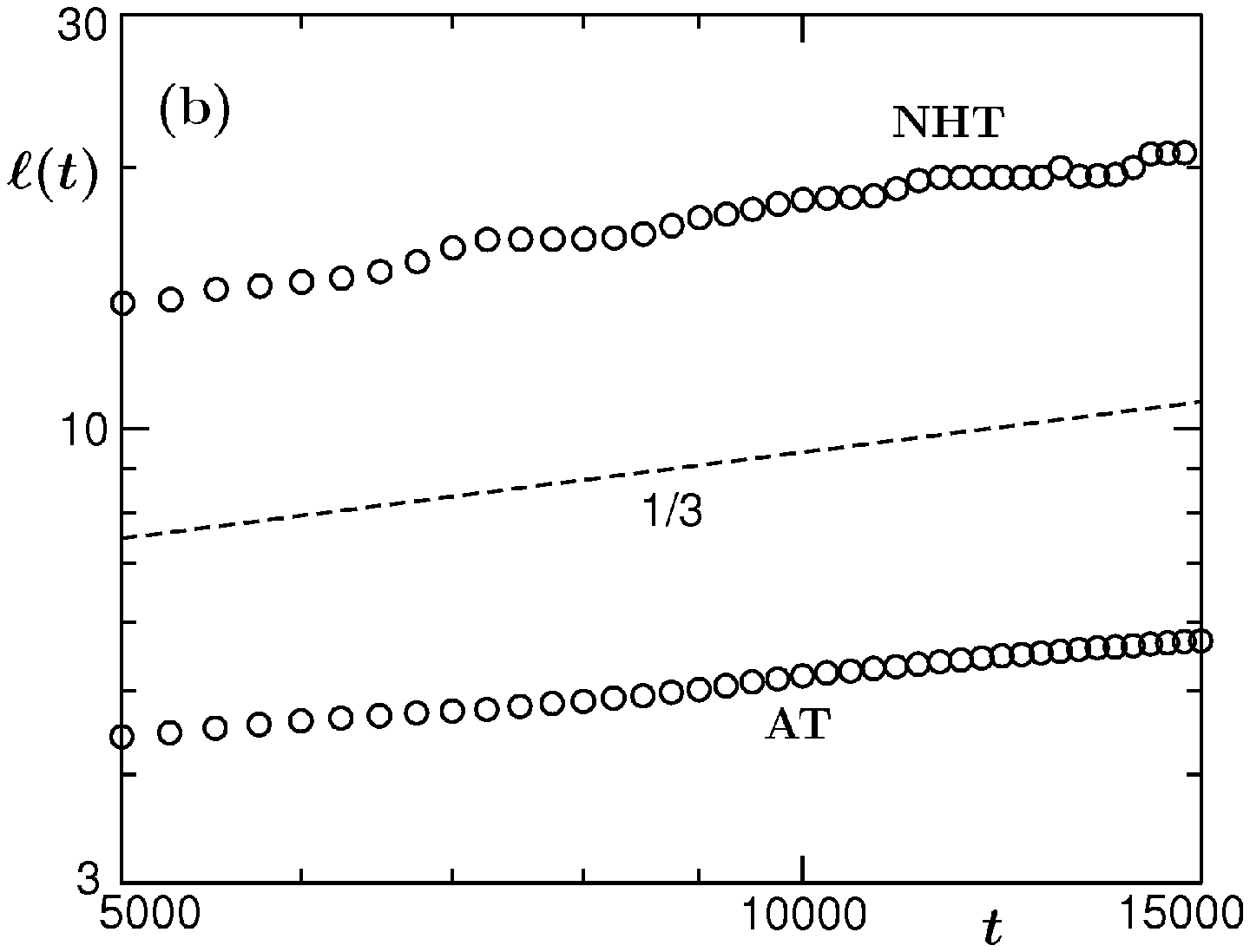}
\caption{\label{fig5}(a) Log-log plot of average domain size, $\ell(t)$, 
as a function of time. The circles correspond to the results obtained 
from the decay of the correlation function. The squares are from direct 
measurements. The parameter values are $\rho=0.05$, $T=0.6$ and $L=100$. 
All results correspond to an averaging over $10$ independent initial 
configurations. Possibilities for various different power-laws are 
indicated. (b) Plots of $\ell(t)$, obtained from direct calculation, vs $t$, 
for both NHT and AT.}
\end{figure}
\par
\hspace{0.2cm} In Fig.\ref{fig4} we depict a typical trajectory of 
a droplet starting from 
$t=1000$ to $t=5000$. This looks reasonably Brownian. One can also try to 
calculate the mean squared displacement from such trajectories. However, 
because of limited number of droplets as well as availability of limited 
time before they collide, it is extremely difficult to obtain data of 
presentable quality. 
\par
\hspace{0.2cm} Having been convinced about the Brownian motion of droplets, 
in Fig.\ref{fig5}(a) we present the plot of $\ell(t)$ vs $t$, on log scale, 
from both types of calculation as described above. While there is overall 
consistency between the two methods, noticeable discrepancy at early time 
is due to the non-scaling behavior of $C(r,t)$ for $t<2000$. Almost 
constant value of $\ell(t)$ uptill approximately $t=100$ is 
indicative of the delayed formation of nucleus of critical size 
as one moves closer to the 
coexistence curve. On the other hand, at late times (when the domain 
densities ``almost" equilibrated) the data are very consistent with the 
predicted BS value $\alpha=1/3$. Note that in this regime we deal with stable 
droplets whose sizes change only after collision. In this context, however, 
calculation of critical nucleus size could be useful, a good discussion of which 
is provided in Ref. \cite{reguera}. Here one may ask the question: how to 
distinguish this from the LS law that also predicts same value for 
$\alpha$? In fact, the same system we have studied via application of an 
Andersen thermostat (AT) \cite{frenkel}. Note that in AT the particles collide 
randomly with the heat reservoir and so stochastic in nature. In 
that situation, the local conservation of momentum is not maintained as 
required in hydrodynamics. In such a case, the domains should grow due to 
diffusion of density leading to the LS value of the exponent. 
This is different from the droplet diffusion in the BS mechanism. 
Indeed we observe that the centre of masses of droplets are static in the 
AT case as expected for LS mechanism. In Fig.\ref{fig5}(b) we presented 
a comparison between the two cases. For the AT, clearly the amplitude of 
growth is much smaller than the NHT. The amplitudes obtained from 
Fig.\ref{fig5}(b) is $A_{BS}/A_{LS}\simeq 3.4$ which differs from the 
theoretical estimate by a factor $\simeq 1.5$ (it becomes closer to $2$ 
if we take $K=4.84$). This discrepancy could be attributed to the 
fact that possibly there is inter-droplet interaction mechanism in 
addition to the BS one.
\par
\hspace{0.2cm} Next we focus on the part of the plot in Fig.\ref{fig5}(a) 
where, immediately after the nucleation of droplets, there is a rapid rise 
of $\ell(t)$. There have been arguments for linear growth in line of  
viscous hydrodynamics, for early time dynamics. This is \cite{tanaka1} 
keeping with the fact that at this early stage, when there is high 
density of droplets, one has nearly interconnected domain structure as 
in case of critical quench. However, our result is more consistent 
with inertial hydrodynamic growth ($\alpha=2/3$), the last scaling 
regime for critical quench. In this latter case one expects a 
competition between growth and break-up of interconnected structures. 
Indeed, in the present case, even though the domains are connected in the 
time regime of discussion, they break up fast due to rapid equilibration 
of density. Nevertheless, we caution the reader that this early time 
result should not be taken seriously due to lack of scaling as seen in 
the inset of Fig.\ref{fig3}.
\par
\hspace{0.2cm} In summary, we studied kinetics of vapor-liquid phase 
separation in a single component Lennard-Jones system. For quenches close 
to the critical density, we observe a percolating structure of vapor and 
liquid domains which grows very rapidly because of the hydrodynamic 
effects. On the other hand, for quenches close to the coexistence density 
(we considered only the vapor branch of the coexistence curve), formation 
and growth of disconnected liquid droplets are observed. Depending upon 
the proximity to the co-existence curve the nucleation of such droplets 
can be significantly delayed. 
\par
\hspace{0.2cm} At late times, the motion and growth of these droplets are 
consistent with the prediction of Brownian diffusion and collision 
mechanics by Binder and Stauffer. Growth in the same system via 
Lifshitz-Slyozov (LS) mechanism has also been studied. The amplitude 
ratio for the BS and LS mechanism is obtained and compared with the 
theoretical predictions. Also, possible reasons for extraordinary fast 
growth, observed before the asymptotic $t^{1/3}$ regime is reached, 
have been pointed out. It will now be interesting to study, among 
other things, the growth dynamics as one continuously changes the 
overall density towards the critical value.
\par
\hspace{0.2cm} The authors acknowledge financial support from the 
Department of Science and Technology, Government of India, via 
Grant No SR/S2/RJN-$13/2009$. SR also acknowledges financial 
support from Council of Scientific and Industrial Research, India.
\par
$*$~das@jncasr.ac.in

\end{document}